\documentclass[runningheads, twocolumn]{aastex63}
\usepackage{mathastext}

\usepackage{newfloat}
\usepackage{natbib}
\usepackage{soul}
\DeclareFloatingEnvironment[name={Supplementary Figure}]{suppfigure}
\DeclareFloatingEnvironment[name={Supplementary Table}]{supptable}

\makeatletter
\def\input@path{{figures/}}
\makeatother

\graphicspath{{figures/}}

\received{}
\revised{}
\accepted{}

\submitjournal{ApJ}

\begin{document}

\title{H$_2$ formation on interstellar grains and the fate of reaction energy}

\correspondingauthor{Stefano Pantaleone}
\email{stefano.pantaleone@univ-grenoble-alpes.fr}

\author[0000-0002-2457-1065]{Stefano Pantaleone}
\altaffiliation{present address: Dipartimento di Chimica and Nanostructured Interfaces and Surfaces (NIS) Centre, Universit\`{a} degli Studi di Torino, via P. Giuria 7, 10125, Torino, Italy.}
\altaffiliation{and}
\altaffiliation{Dipartimento di Chimica, Biologia e Biotecnologie, Universit\`{a} di Perugia, Via Elce di Sotto 8, 06123 Perugia, Italy}
\affiliation{Univ. Grenoble Alpes, CNRS, Institut de Plan\'{e}tologie et d'Astrophysique de Grenoble (IPAG), 38000 Grenoble, France}

\author[0000-0002-2147-7735]{Joan Enrique-Romero}
\affiliation{Univ. Grenoble Alpes, CNRS, Institut de Plan\'{e}tologie et d'Astrophysique de Grenoble (IPAG), 38000 Grenoble, France}
\affiliation{Departament de Qu\'{i}mica, Universitat Aut\`{o}noma de Barcelona, Bellaterra, 08193, Catalonia, Spain}

\author[0000-0001-9664-6292]{Cecilia Ceccarelli}
\affiliation{Univ. Grenoble Alpes, CNRS, Institut de Plan\'{e}tologie et d'Astrophysique de Grenoble (IPAG), 38000 Grenoble, France}

\author[0000-0001-7819-7657]{Stefano Ferrero}
\affiliation{Departament de Qu\'{i}mica, Universitat Aut\`{o}noma de Barcelona, Bellaterra, 08193, Catalonia, Spain}

\author[0000-0001-5121-5683]{Nadia Balucani}
\affiliation{Univ. Grenoble Alpes, CNRS, Institut de Plan\'{e}tologie et d'Astrophysique de Grenoble (IPAG), 38000 Grenoble, France}
\affiliation{Dipartimento di Chimica, Biologia e Biotecnologie, Universit\`{a} di Perugia, Via Elce di Sotto 8, 06123 Perugia, Italy}
\affiliation{Osservatorio Astrofisico di Arcetri, Largo E. Fermi 5, 50125 Firenze, Italy}

\author[0000-0002-9637-4554]{Albert Rimola}
\affiliation{Departament de Qu\'{i}mica, Universitat Aut\`{o}noma de Barcelona, Bellaterra, 08193, Catalonia, Spain}

\author[0000-0001-8886-9832]{Piero Ugliengo}
\affiliation{Dipartimento di Chimica and Nanostructured Interfaces and Surfaces (NIS) Centre, Universit\`{a} degli Studi di Torino, via P. Giuria 7, 10125, Torino, Italy.}

\begin{abstract}
Molecular hydrogen is the most abundant molecular species in the Universe. 
While no doubts exist that it is mainly formed on the interstellar dust grain surfaces, many details of this process remain poorly known.
In this work, we focus on the fate of the energy released by the H$_2$ formation on the dust icy mantles, how it is partitioned between the substrate and the newly formed H$_2$, a process that has a profound impact on the interstellar medium. 
We carried out state-of-art \textit{ab-initio} molecular dynamics simulations of H$_2$ formation on periodic crystalline and amorphous ice surface models.
Our calculations show that up to two thirds of the energy liberated in the reaction ($\sim$300 kJ/mol $\sim$3.1 eV) is absorbed by the ice in less than 1 ps.
The remaining energy ($\sim$140 kJ/mol $\sim$1.5 eV) is kept by the newly born H$_2$.
Since it is ten times larger than the H$_2$ binding energy on the ice, the new H$_2$ molecule will eventually be released into the gas-phase.
The ice water molecules within $\sim$4 \AA ~from the reaction site acquire enough energy, between 3 and 14 kJ/mol (360--1560 K), to potentially liberate other frozen H$_2$ and, perhaps, frozen CO molecules.
If confirmed, the latter process would solve the long standing conundrum of the presence of gaseous CO in molecular clouds.
Finally, the vibrational state of the newly formed H$_2$ drops from highly excited states ($\nu = 6$) to low ($\nu \leq 2$) vibrational levels in a timescale of the order of ps. 

\end{abstract}


\section{Introduction}\label{sec:intro}

Molecular hydrogen is the most abundant molecule in the Universe.
Its formation is also the first step of the interstellar chemistry and, therefore, a fundamental reaction.
In molecular clouds, H$_2$ is mainly formed via the H + H association reaction on the interstellar dust grain surfaces, which act as a third body capable to partially absorb the energy ($\sim$440 kJ/mol $\sim$4.56 eV) released by the chemical reaction and, consequently, stabilise the newly formed H$_2$ molecule \citep[e.g.][]{van1946,hollenbach1970,hollenbach1971,h_diff1}. 
Although no doubts exist on the occurrence of this process, many specific details remain poorly known \citep[see e.g.][]{vidali2013,h2_form}.

Here, we focus on the fate of the energy released by the reaction, which has been source of debate for decades. Specifically: how much of the reaction energy is absorbed by the dust grain and in what timescale? What fraction of the chemical energy does go into the kinetic energy of the newly formed H$_2$? Is this energy large enough as H$_2$ breaks the interaction with the surface and leaves into the gas-phase? How much energy does the H$_2$ molecule possess when it leaves the grain surface and in what ro-vibrational state? And, finally, is the energy transmitted to the dust grain enough for locally warming it up and make adjacent frozen molecules sublimate? These points have a profound impact on several aspects of astrochemistry and, more generally, the physics and chemistry of the interstellar medium (ISM) \citep{DW1993}. 

Answers to the above questions largely depend on the nature of the substrate, namely the specific dust grain surface or, in practice, the interstellar environment where H$_2$ forms. Here we focus on cold ($\sim 10$ K) molecular clouds. In these environments, the grain refractory cores are coated by water-dominated icy mantles either in polycrystalline (PCI) or amorphous (ASW for amorphous solid water) phases, the latter dominating over the former \citep{smith1988,whittet1993,Boogert2015}.

Unfortunately, laboratory experiments cannot entirely reproduce the interstellar conditions so that, while they can suggest routes and processes, they cannot provide definitive answers to the above-exposed questions \citep[see e.g.][]{vidali2013}. For example, obtaining experimentally how the H$_2$ formation nascent energy is partitioned is very difficult \citep{roser2003,hornekaer2003,hama2012,watanabe2010}.
Computational chemistry methods can be regarded as a complementary tool and, sometimes, as a unique alternative to laboratory experiments. So far, a limited number of studies devoted to the energy dissipation of the H$_2$ formation reaction on icy grains has appeared in the literature \citep{takahashi1999c,takahashi2001,herbst2006monte}.

Here, we present a new theoretical study on the dissipation of the energy released by the $H + H \longrightarrow H_2$ reaction on water ice using {\it ab-initio} molecular dynamics simulations (AIMDs). We simulated the reaction adopting a Langmuir-Hinshelwood (LH) mechanism (\textit{i.e.} both reactants are adsorbed one the surface in neighboring sites) both on amorphous and crystalline periodic models of interstellar icy mantles.



\section{Computational details} \label{sec:comp-details}

\subsection{Methods}\label{subsec:methods}
All our calculations were carried out with the CP2K package \citep{cp2k1}. Core electrons were described with the Goedecker-Teter-Hutter pseudopotentials \citep{pseudo1} and valence ones with a mixed Gaussian and Plane Wave (GPW) approach \citep{gpw}. The DFT (Density Functional Theory) PBE (Perdew-Burke-Ernzerhof) method was adopted \citep{pbe}, combined with a triple-$\zeta$ basis set for valence electrons with a single polarization function (TZVP), and the \textit{a posteriori} D3 Grimme correction to account for dispersion forces \citep{grimme1,grimme2}.
The plane-wave cutoff was set to 600 Ry. Ice surfaces were thermalized (see sec. Ice models), and during the geometry optimization only the reacting H atoms (those forming H$_2$) were allowed to move, keeping the water molecules fixed at their thermalized positions. All calculations were carried out within the unrestricted formalism as we deal with open-shell systems (see SI for more details). The binding energy (BE) of H$_2$ was calculated as:
\begin{equation}
BE_{H_2}=E_{CPLX}-(E_{Ice} + E_{H_2})
\end{equation}
\noindent where $E_{CPLX}$ is the energy of the H$_2$/Ice system, $E_{Ice}$ that of the bare ice surface, and $E_{H_2}$ the energy of the H$_2$ alone.

AIMDs were carried out within the NVE (N = number of particles, V = volume, E = energy) ensemble, where the total energy V$_{TOT}$ (\textit{i.e.} potential + kinetic) is conserved. The evolution of the system was followed for 5 ps for the crystalline model and for 1 ps for the amorphous model, using time-steps of 0.2 fs.

\subsection{Ice models}\label{subsec:ices}
For the crystalline ice model, a periodic slab cut from the hexagonal ice bulk structure was used. The periodic cell parameters defining the system are: $\textit{a} = 26.318$ \r{A} and $\textit{b} = 28.330$ \r{A}, while the slab thickness is $\sim21$ \r{A} (corresponding to seven layers), for a total of 576 water molecules. The \textit{c} parameter of the simulation box, \textit{i.e.} the non-periodic one, was set to 50 \r{A} to avoid interactions among the fictitious slab replicas. The size of the slab model was chosen to avoid non-physical temperature increase due to the extremely large exothermic reaction \citep[][]{pantaleone2020chemical}.


The amorphous model was obtained by performing a classical molecular dynamics (MD) simulation on the crystalline structure. The MD was carried out with the TIP3P force-field \citep{tip3p} for 200 ps (with a timestep of 0.5 fs) at 300 K within the NVT (N = number of particles, V = volume, T = Temperature) ensemble. A second NVT simulation was performed at 10 K, to cool down the system at the temperature of cold molecular clouds. Finally, a geometry optimization and another NVT-MD simulation at 10 K were run using PBE, to recover the potential energy surface at the same theory level as described in the previous section.


\subsection{\texorpdfstring{H$_2$ vibrational state}{H2 vibrational state}}\label{subsec:comp_vib}

To calculate the H$_2$ vibrational state during the AIMDs the anharmonic oscillator model was employed. As a first step, the PES (potential energy surface) of the H$_2$ isolated molecule was explored by performing a rigid scan of the H--H distance, ranging from 0.3 to 3.0 \r{A} with a step of 0.01 \r{A}. Computed data were fitted with the Morse equation to obtain the force constant of the oscillator, its dissociation energy and, hence, the vibrational levels of H$_2$. The H$_2$ turning points calculated with our model were compared with those of the AIMD simulations and the vibrational level was assigned at each H$_2$ oscillation during the AIMDs.


\section{Results}\label{sec:res}

\subsection{Reactants plus product positions in the ice models}\label{subsec:res-ices}

\begin{figure}
\centering
\includegraphics[width=\columnwidth]{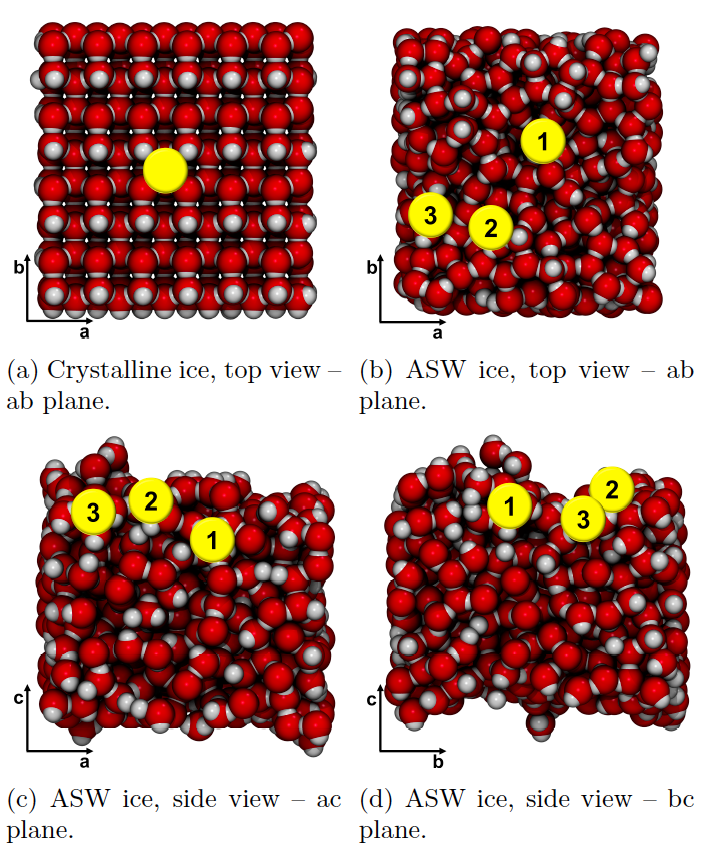}
    \caption{Top view of the crystalline (a), and top (b) and side (c and d) views of the ASW ice models. 
    The yellow circles represent the centers of mass of the two H atoms starting positions. 
    In the case of ASW, the numbers mark the positions (Pos1, Pos2 and Pos3) discussed in the text.
    The initial H to H distance is about 4 \AA\, (see Supplementary Fig. \ref{h2_ene}).
    O atoms are in red and H atoms of the ice are in white.}
\label{fig:start-pos}
\end{figure}

\begin{figure}
\centering
\includegraphics[width=\columnwidth]{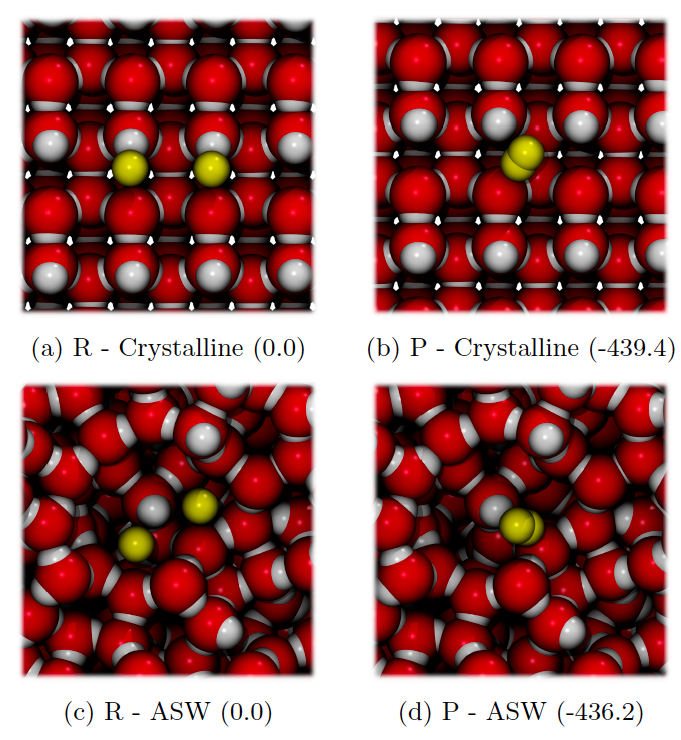}
    \caption{PBE-D3/TZVP optimized geometries of reactants (left panels) and product (right panels) of the H$_2$ reaction formation on the crystalline (top) and ASW Pos1 (bottom) ice models. 
    The numbers in parenthesis correspond to the relative energy in kJ/mol with respect to the reactants. 
    The H atoms involved in the H + H $\longrightarrow$ H$_2$ reaction are in yellow, those belonging to surface water molecules are in white, and O atoms are in red.}
    \label{fig:ices}
\end{figure}

\begin{table*}[bt]
    \caption{Results of the AIMD for the H$_2$ formation on the crystalline (first row) and the ASW three positions (Fig. \ref{fig:start-pos}) (bottom three rows) ices.
    Second and third columns report the fraction of kinetic energy of the ice and the newly formed H$_2$, respectively, at the end of the simulation (5 and 1 ps for crystalline and ASW ices, respectively).
    Fourth column reports the H$_2$ binding energy (in kJ/mol) at the reaction site.
    Fifth and sixth columns report the vibrational state averaged over 0.2--1 ps and 1--5 ps, respectively (the latter can be computed only for the crystalline case).
    Last column reports the peak energy (in kJ/mol) of an ice-molecule neighbor (i.e. at $\sim 3$ \AA) to where the reaction occurs.
    }
    \centering
    \begin{tabular}{ccccccc}
    \hline
     Model &  $\frac{K_{ice}}{K_{ice}+K_{H_2}}$ & $\frac{K_{H_2}}{K_{ice}+K_{H_2}}$ & BE$_{H_2}$ & $\nu^{0.2--1}$ & $\nu^{1--5}$ & E$_{peak}$ \\
           & & & [kJ/mol] & & & [kJ/mol] \\
     \hline
    Crys. & 0.45 & 0.55  & 9.0  & 4--5 & 1--2 & 12.0 \\
    Pos1  & 0.65 & 0.35  & 9.7  & 1--2   & -- & 43.0\\
    Pos2  & 0.46 & 0.54 & 10.6  & 6      & -- & 7.0\\
    Pos3  & 0.47 & 0.53 & 12.4  & 5      & -- & 7.0\\
    \hline
    \end{tabular}
    \label{tab:results}
\end{table*}

Figure \ref{fig:start-pos} shows the structure of the used crystalline and ASW ice models as well as the starting positions of the center of mass of the two adsorbed H atoms before the reaction.The starting H--H distance, as well as its evolution during the AIMDs is shown in Supplementary Figure \ref{h2_ene}.
The ASW model presents a rugged morphology (e.g. holes, cavities, channels) in comparison to the crystalline one and, therefore, we carried out simulations in three positions, roughly representing different possible situations in terms of energetics and surface morphology. 
Position marked as Pos1 is the deepest one, within a cavity, while positions Pos2 and Pos3 are in the outermost parts of the surface. 

As a first step, we optimized the geometries of reactants (the two H atoms) and product (the H$_2$ molecule) on the water ice surface in order to obtain the potential energy surface of the reaction. 
No search for transition state structures has been performed because AIMDs indicated this reaction is barrierless at 10 K. 
That is, the starting kinetic energy of the two H atoms provided by the 10 K is high enough to overcome the eventually small potential energy barrier. 
The total energy to be dissipated is around 435-440 kJ/mol, depending on the surface and starting position. 

Figure \ref{fig:ices} provides a view of the two H atoms starting positions and H$_2$ position immediately after the reaction, on the crystalline and ASW (Pos1 as an example) models, respectively.

%
\subsection{Molecular hydrogen desorption}\label{subsec:h2des}

AIMDs results are summarised in Table \ref{tab:results} and shown in Fig. \ref{fig:tot_ene}.
First, at least half of the kinetic energy is absorbed by the ice while the remaining one is kept by the newly formed H$_2$ molecule ($\sim$50--35\%).
These numbers are similar for both for the crystalline and ASW ice positions, suggesting they do no depend much on the surface structural details of the ice.

Second, after the H--H bond formation, the newly formed H$_2$ molecule diffuses over the surface, as it keeps a large kinetic energy, in both crystalline and ASW ice models.
However, despite the energetics of the two processes are similar (Figs. \ref{fig:tot_ene} left panels) the H$_2$ diffusion (Figs. \ref{fig:tot_ene} right panels) is different. 

On the crystalline surface, the diffusion of the newly formed H$_2$ over the ice surface is constrained in a specific direction, along the a-axis, parallel to the ice surface, whereas in the b-axis and c-axis directions the starting position does not change.
This is because the crystalline model has alternated and opposite electrostatic potentials along the b-axis direction (see Supplementary Fig. \ref{esp}, which constrains the H$_2$ diffusion to the perpendicular a-axis, within a channel-like structure created in between the two alternated potential regions (see Figs. S2a and S2c).
In contrast, on the ASW ice model, H$_2$ diffuses over all the three directions, also along the c-axis which corresponds to direction perpendicular to the ice surface. 
However, the movement is not the same in the three studied positions.
In Pos1, H$_2$ achieves a maximum height of $\sim$10 \r{A} over the surface in a timescale of 0.4 ps and does not come back (Fig. \ref{fig:tot_ene}-d), this way leaving definitely the surface.
In Pos2, on the contrary, H$_2$ slides and hops on the surface within the 1 ps of the simulation (Fig. \ref{fig:tot_ene}-f).
A similar behavior is observed for Pos3, but with wider hops and jumps (Fig. \ref{fig:tot_ene}-h).
However, and this is the key point of the simulations, the kinetic energy remains much larger ($\sim ~150$ kJ/mol) than the binding energy ($\sim 10$ kJ/mol) in all cases: therefore, the nascent H$_2$ is likely doomed to leave the ice surface.

\subsection{Energy dissipation towards the ice water molecules}\label{subsec:e-dissi}
Fig. \ref{fig:wat_kin} shows how the energy absorbed by the ice is first transmitted from reaction site to a neighbor water(i.e. at $\sim$4 \AA) in $\sim$200--800 fs and, with time, to shells of water molecules with increasing distances.
The neighbor water molecule acquires from $\sim$3 to $\sim$14 kJ/mol and stays with that energy for more than about 100--200 fs before the energy is dissipated towards more external water molecules.
These timescales are in excellent agreement with previous studies of the sound speed in ices \citep[see e.g.][]{ruocco_equivalence_1996}.

\subsection{\texorpdfstring{H$_2$ vibrational state}{H2 vibrational state}}\label{subsec:h2-vib}

The initial state of the H$_2$ formed on the crystalline ice is $\nu$ = 4--5 (obtained averaging over 0.2--1 ps) and then it decreases to $\nu$ = 1--2 (1--5 ps average) because of the H$_2$ energy dissipation.
Conversely, at Pos1 of the ASW model, H$_2$ is formed with a low vibrational state ($\nu$ = 1--2, 0.2--1 ps average).

\begin{figure*}
    \centering
    \includegraphics[width=0.9\textwidth]{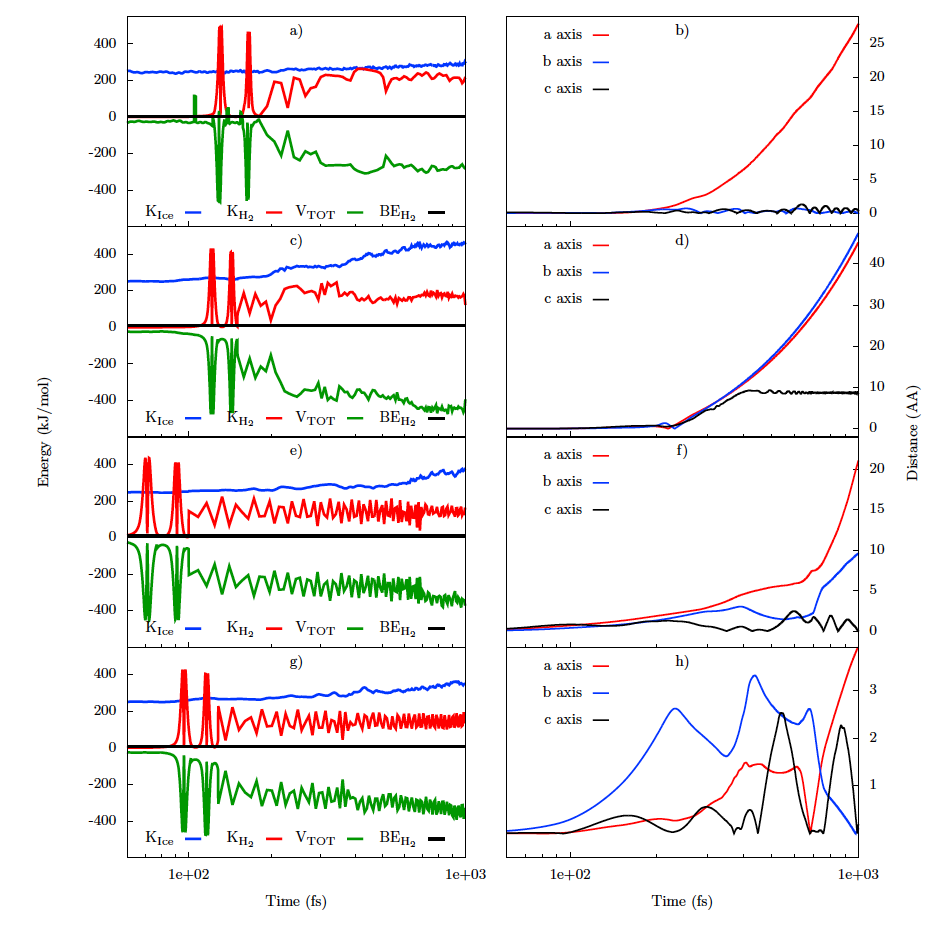}
    \caption{Results of the AIMDs for the crystalline (top panels) and ASW Pos1, Pos2, Pos3 (panels on the second, third and bottom rows) models, respectively.
    \textit{Left panels:} Evolution over time (in fs) of the most relevant energetic components (in kJ/mol) of the H$_2$/ice system for the crystalline (panel a) and ASW Pos1, Pos2 and Pos3 (panel c, e and g) models, respectively. 
    V$_{TOT}$ is the total potential energy (green lines), K$_{H_2}$ and K$_{Ice}$ are the kinetic energies of H$_2$ (red lines) and ice (blue lines), respectively, while BE$_{H_2}$ (10 kJ/mol = 1200 K) is the binding energy of the H$_2$ (black lines).
    \textit{Right panels:} Diffusion of the center of mass of the H$_2$ molecule split into the three cartesian components: c-axis is the direction perpendicular to the ice surface, while a-axis and b-axis are along the ice surface (\S ~\ref{sec:comp-details}). }
    \label{fig:tot_ene}
\end{figure*}%

For the other two positions (Pos2 and Pos3), the vibrational state of the newly formed H$_2$ molecule lies in highly excited states, $\nu$ = 6 and $\nu$ = 5, respectively.
However, our simulations stop at 1 ps and, likely, as on the crystalline ice, the vibrational state of the newly formed H$_2$ will lower to $\nu$ = 1--2.
The difference in the behavior between the Pos1 and Pos2/3 is probably due to a transient chemical bond between one of the reacting H and one O atom of a neighboring water molecule (see Supplementary Figure \ref{kin_supp}.


\section{Discussion}\label{sec:disc}

\subsection{H\texorpdfstring{$_2$}{2} and other molecules desorption}
While there were no doubts that, once formed on the icy surfaces, H$_2$ molecules will be released into the gas-phase, our new computations provide a quantitative and atomistic-view support to this theory.
The newly formed H$_2$ molecules possess an energy much larger than their binding energy, by more than a factor ten, so that very likely H$_2$ will be injected into the gas-phase.
Even in the worst case of crystalline ice, where the simulations show H$_2$ sliding over the perfect ice surface, the newly formed H$_2$ will sooner or later stumble on an imperfection of the ice so that its trajectory would be deviated and the molecule escape into the gas. To have an estimation of the timescale of this phenomenon one could consider the timescale for the newly H$_2$ to scan the entire surface of an interstellar grain. Assuming its radius equal to 0.1 $\mu$m and each icy site size equal to 3 \r{A}, the number of sites to scan are about 10$^5$.
A first estimate of the time to scan all the sites can be directly obtained by our simulations by considering that about 15 sites of our crystalline ice are covered in less 1 ps: therefore, to scan the 10$^5$ it will take less than 10 ns.
However, this is strictly true if the H$_2$ molecule does not loose energy in other minor impacts, so $\sim 10$ ns can be considered a lower limit.
The upper limit can be computed by assuming that no residual energy is left to the H$_2$ molecule except the thermal one and, in this case, the velocity to scan is given by the hopping rate and the timescale for scanning the entire grain surface becomes: 

\begin{equation}
t_{scan}=N_{s}\nu_0^{-1}e^{\frac{E_d}{k_BT_{dust}}}
\end{equation}
\noindent where $\nu_0$ is about 10$^{12}$ s$^{-1}$ and E$_d$ is assumed to be 0.3 times the binding energy (about 400 K). Inserting the numbers, a rough estimate of the time that H$_2$ takes to leave a typical interstellar grain is $\leq 1000$ yr. 
In laboratory analogues of interstellar grains, the timescale would be even larger, and, consequently, not observable.
In conclusion, the newly formed H$_2$ will leave the surface in a timescale between a few ns to a max of about 1000 yr, which is still a very short lifetime with respect to the cloud life.

We want to highlight that the choice of using a proton order ice is just a matter of convenience to test our simulations, before going to the more realistic case of the amorphous surface. On a more realistic proton disordered crystalline ice we expect a more anisotropic behavior of the H$_2$ molecule, and, as a consequence, a faster H$_2$ desorption.

\begin{figure*}
\centering
\includegraphics[width=0.9\textwidth]{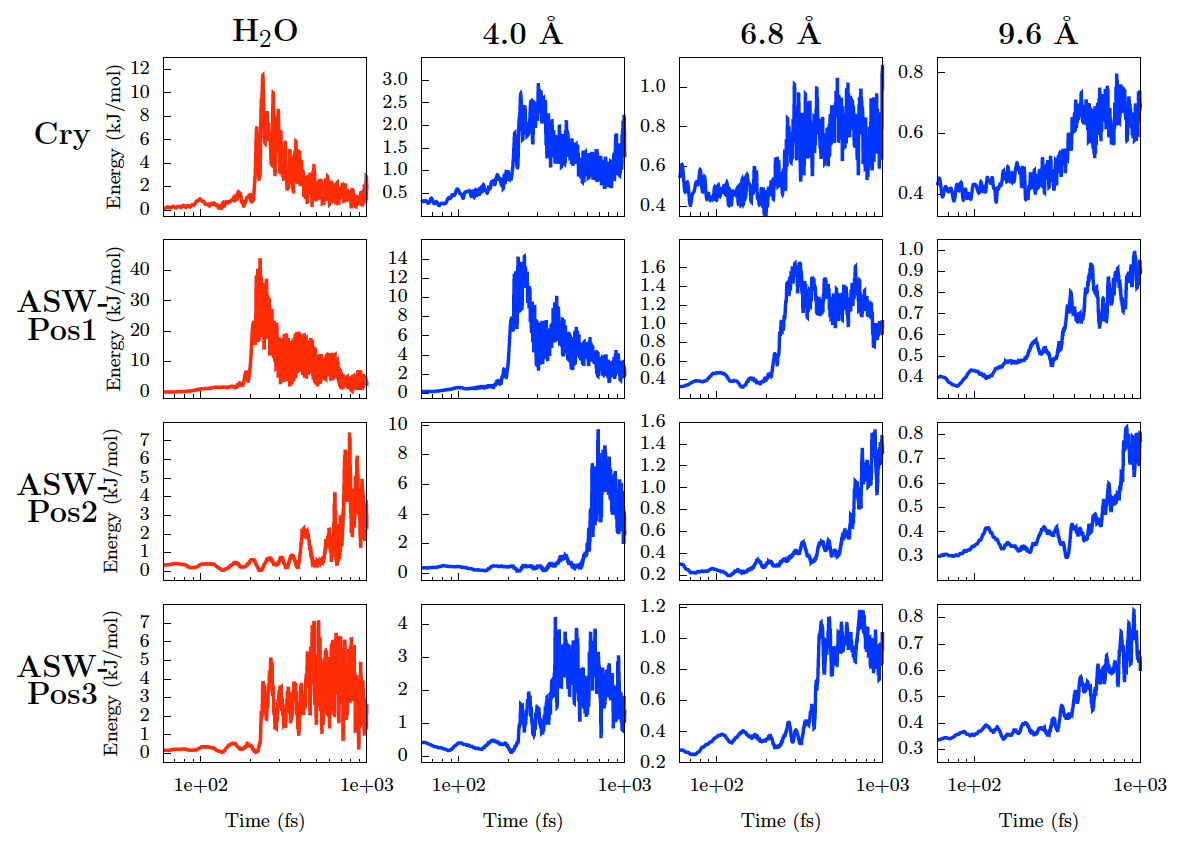}
\caption{Kinetic energy (in kJ/mol) acquired by the water ice molecules as a function of time for the crystalline ({\it top first row panels}) and ASW Pos1-3 ({\it lower panels}) ices, as marked by the left-hand labels. 
{\it Left first column panels}: Energy acquired by a single neighboring water ice molecule (i.e. at $\sim3$ \AA) from the H$_2$ reaction site. 
{\it Other column panels}: Energy  acquired by the ice surface divided in concentric shells (normalized by the number of water molecules per shell), centered at the reaction site. 
The labels on the top mark the shells radii.}
\label{fig:wat_kin}
\end{figure*}

Our computations also show that the ice absorbs a significant fraction of the reaction energy, 45--65\% (Tab. \ref{tab:results}).
This energy propagates like a wave from the point where the H + H reaction occurs \citep[see Fig. \ref{fig:wat_kin} and also ][]{pantaleone2020chemical}.
In ASW ice, a water molecule close to reaction site acquires 7--43 kJ/mol, (840--5160 K), for more than 100-200 fs.
Within the first shell of radius 4 \AA, water molecules acquire energies from 3 to 14 kJ/mol (360 to 1680 K, average value counting all the water molecules within the first shell). and farther away the energy acquired by the ice-water molecules decreases to less than 1.6 kJ/mol.
The energy acquired by the water ice molecules within a radius of 4 \AA~ from the reaction site could, potentially, be enough to release into the gas-phase any molecule whose binding energy is lower than 3--14 kJ/mol.
This could be the case of another H$_2$ molecule frozen on the ice, a possibility predicted by some astrochemical models \citep[see e.g.][]{hincelin2015new}.
Indeed, since the binding energy of H$_2$ is less than $\sim$5 kJ/mol \citep[e.g.][]{vidali2013,Ferrero2020, Molpeceres_Kastner_2020}, several H$_2$ molecules could be kicked into the gas-phase.

More interesting is the case of CO, the most abundant molecule after H$_2$ in galactic cold molecular clouds. 
It has a binding energy between 7 and 16 kJ/mol \citep[e.g.][]{He2016,Ferrero2020} so that frozen CO could potentially be liberated into the gas-phase by the formation of H$_2$.
It has been long recognized that, in absence of a non-thermal desorption process, CO in molecular clouds should be entirely frozen onto the interstellar ices in $\sim3\times 10^5$ yr \citep[see e.g.][]{leger1983does}.
Several mechanisms have been proposed to explain its presence in the gas \citep[see e.g.][]{leger1985desorption,schutte1991explosive,DW1993,shen2004cosmic,ivlev2015impulsive}. 
They are all based on the idea that the grain is locally heated and that the frozen molecules are statistically desorbed on a time scale of a few thousands seconds \citep[see e.g.][]{hasegawa_new_1993,herbst2006monte}, except when faster photodesorption processes are involved.
In particular, Duley \& Williams \citep{DW1993} and Takahashi \& Williams \citep{takahashi2000chemically} focused on the CO desorption caused by the H$_2$ formation.
Using the results of classical MD simulations by Takahashi and colleagues \citep{takahashi1999a,takahashi1999b,takahashi2000chemically} and a statistical approach, these authors found that the surface within a radius of $\sim$4 \AA~ from the reaction site is heated up to $\leq$40 K for a too short time ($\sim 10$ fs) to allow CO to sublimate.
However, despite the Takahashi's simulations are impressive considering that they were carried out more than 20 years ago, old force fields (like the TIP3P used by Takahashi et al.) have limitations in describing the actual intermolecular interactions and the energy transfer from one molecule to another, compared to our simulations treated at quantum mechanical level. 

Our AIMDs show that the water molecules within a radius of 4 \AA~ from the reaction site can be excited from 3 to 14 kJ/mol (depending on the position) for more than 100 fs  (Fig. \ref{fig:wat_kin}). 
Using the range of values in the literature for the CO binding energies  \citep[7--16 kJ/mol: ][]{He2016,Ferrero2020}, and the usual Eyring equation to estimate the half-life time of CO desorption, we obtain that a CO molecule within a radius of 4 \AA~ from Pos1 would desorb in 33--62 fs, and 69--185 fs from Pos2.
On the contrary, CO molecules close to Pos3 and on crystalline ice would not desorb.
Therefore, based on this rough argument, the formation of H$_2$ can potentially desorb frozen CO molecules.
Whether this hypothesis is realistic or not, it depends on (i) how many sites of the AWS ice water molecules are excited as in Pos1 and Pos2, (ii) the ratio of  H$_2$ formation rate with respect to the CO freezing rate ($R=\frac{v(H_2)}{v(CO)}$), and (iii) the probability that CO and H$_2$ are adjacent.

While a more realistic model for the ASW ice is needed to estimate the first quantity (i), one can estimate the second one (ii) as follows.
At steady state, assuming that the H + H reaction has efficiency 1 \citep[e.g.][]{vidali2013}, the ratio $R$ of the H$_2$ formation rate with respect to the CO freezing rate is equal to the ratio between the rate of H atoms over CO molecules landing on the grain surface, divided by 2 (because two H atoms are needed for the H$_2$ formation).
Considering an average Milky Way molecular cloud with H$_2$ density of $10^3-10^4$ cm$^{-3}$, temperature 10 K, a gaseous (undepleted) CO abundance with respect to H$_2$ equal to $2 \times 10^{-4}$, and a cosmic ray ionisation rate of $3\times10^{-17}$ s$^{-1}$, one obtains $\rm{n_H}\sim 2$ cm$^{-3}$ and $R \sim 26-2.6$. That is, the H$_2$ formation rate dominates over the CO freezing one.
Therefore, frozen CO molecules can potentially be desorbed by the energy released by the H$_2$ formation on the icy grain surfaces.
This process, if confirmed, might naturally explain the presence of gaseous CO in not too dense (n$_H\lesssim10^4$ cm$^{-3}$) molecular clouds and solve a decades long mystery.
However, to put this affirmation on a solid ground, specific AIMDs showing the effective sublimation of the CO molecule as well as larger ASW ice models and dedicated astrochemical modeling simulations that include this effect are mandatory. 
They will be the focus of forthcoming works.

\subsection{H\texorpdfstring{$_2$}{2} vibrational state}
Previous experimental and computational works on graphite surfaces show that after its formation, the H$_2$ molecule populates vibrational levels around $\nu$ = 3--4 (\cite{latimer2008,islam2010,casolo2013}).
On crystalline and amorphous ice surfaces only a few studies were carried out. 
Based on classical MD simulations, \cite{takahashi1999c,takahashi2001} predicted that H$_2$ formed on amorphous ice would be vibrationally excited to $\nu$ = 7--8.
In contrast, in some laboratory experiments \cite{roser2003,hornekaer2003,hama2012}, the authors did not detect such excited states. 
Our new simulations confirm both findings: depending on the site where the formation occurs, H$_2$ can have large (up to 6) or low (1--2) $\nu$ (Tab. \ref{tab:results}).
For example, the case of ASW Pos1, which is the one in a cavity, shows the lowest $\nu$.
In the ISM, the vibrational state of the newly formed H$_2$ molecules will depend on the probability of H$_2$ leaving the grain surface before being "thermalised" by collisions with ice water molecules, as also found and discussed in \cite{watanabe2010}.

The vibrational state of the newly formed H$_2$ molecule, \textit{i.e.} how much the molecule is vibrationally excited, has important consequences on two major astrophysical aspects:
(i) to help gas-phase reactions with high activation energy barriers, some of which are the starting points of the entire chemistry in molecular clouds \cite{agundez2010};
(ii) the possibility to detect newly formed H$_2$ with near-future JWST (James Webb Space Telescope) observations and, consequently, measure the rate of H$_2$ formation in molecular clouds and put constraints to theories.

\section*{Summary}\label{sec:concl}
We studied the energy dissipation of the $H_2$ formation reaction on both crystalline and amorphous (ASW) ice models by means of state-of-art \textit{ab-initio} molecular dynamics simulations (AIMDs). 
In the ASW ice, we explored three formation sites meant to represent different situations in terms of energetics and surface morphology: one is in a cavity (Pos1) and two in the outermost parts of the surface (Pos2 and Pos3).

In all simulations, we found that about at least 30\% of the reaction energy ($\sim$440 kJ/mol) is acquired by the newly formed H$_2$, namely more than ten times the H$_2$ binding energy ($\sim$10 kJ/mol), so that H$_2$ is likely doomed to leave the ice and to be injected into the gas.
The remaining two thirds of the reaction energy are absorbed by the water ice in less than 1 ps.
The water molecules nearby to the reaction site have energy peaks of 7--43 kJ/mol for more than 100--200 fs, while those within a 4.0 \AA ~radius of 3--14 kJ/mol. 

We showed that it is in principle possible that frozen CO molecules close to the H$_2$ formation site sublimate. 
If confirmed, this will be a simple explanation to the decades-long conundrum of why gaseous CO is present in cold molecular clouds.
In order to quantify this effect new focused AIMDs adopting larger ASW ice models and dedicated astrochemical modeling will be necessary.

Finally, the nascent H$_2$ molecules have large ($\nu = 6$) vibrational states in the first ps and later decay to 1--2.
This high vibrational state could help reactions with an activation barrier involving H$_2$ to occur also in cold gas and be observable by JWST.

\acknowledgments
The authors acknowledge funding from European Union’s Horizon 2020 research and innovation program,  the European Research Council (ERC) Project “the Dawn of Organic Chemistry", grant agreement No 741002,  the European Research Council (ERC) Project "Quantum Chemistry on Interstellar Grains", grant agreement No 865657, and the Marie Skłodowska-Curie  project ``Astro-Chemical Origins” (ACO), grant agreement No 811312.
AR is indebted to the ``Ram{\'o}n y Cajal" program. MINECO (project CTQ2017-89132-P) and DIUE (project 2017SGR1323) are acknowledged.
BSC-MN and OCCIGEN HPCs are kindly acknowledged for the generous allowance of super-computing time through the QS-2019-2-0028 and 2019-A0060810797 projects, respectively.
SP, NB and PU acknowledge the Italian Space Agency for co-funding the Life in Space Project (ASI N. 2019-3-U.O

\bibliography{references}{}
\bibliographystyle{aasjournal}


\appendix \label{appendix}

\section{Computational details}\label{app-comp-details}

Proper reactants spin localisation is achieved by relying on the unrestricted-DFT broken (spin) symmetry approach. 
This is mandatory when dealing with diradical systems in singlet electronic state. 
In the H + H specific case, despite the total spin moment on the global system is 0, the unpaired electrons on the two H atoms are expected to have spin $\alpha$ (\textit{i.e.} +1) and $\beta$ (\textit{i.e.} -1) each.

In order to check the quality of our results, in Table \ref{h2_ene} a comparison between the H$_2$ formation energy in the gas phase calculated at PBE and CCSD(T) levels is presented Moreover, also the interaction between H$_2$ and one H$_2$O molecule was evaluated at the same levels of theory, giving results in good agreement. Moreover, also the interaction between H$_2$ and one H$_2$O molecule was evaluated at the same levels of theory, giving results in good agreement.
The difference between the two methodologies is acceptable, considering the energy released by the reaction.

As regards the AIMD simulations, we run an equilibration AIMD in the NVT ensemble (using the CSVR thermostat, with a time constant of 20 femtoseconds) at 10 K for 1 ps (with a time step of 1 fs) only for the bare ice surface. 
This ensures an initial thermally equilibrated ice. 
Then the velocities of the equilibrated ice surfaces are used as input for the NVE production runs. 
Those velocities, correspond to the initial kinetic energy of the ice, as presented in Figure \ref{fig:tot_ene}. 
The H velocities, instead, were manually set to favor the H--H bond formation respecting the 10 K limitation.

\begin{supptable}[h!]
\caption{Adsorption and reaction (H + H $\longrightarrow$ H$_2$) energy data.}
\begin{center}
\begin{tabular}{lc}
\hline
\hline
 & Reaction energy (kJ mol$^{-1}$)\\
\hline
Crystalline	&	-438.9	\\
Pos1	&	-436.2	\\
Pos2	&	-435.9	\\
Pos3	&	-440.3	\\
H$_2$ CP2K$^a$	&	-440.6	\\
H$_2$ Gaussian$^b$	&	-457.4	\\
\hline
& Binding energy (kJ mol$^{-1}$)\\
\hline
CP2K$^a$ & 4.1\\
Gaussian$^b$ & 2.6\\
\hline
\hline
\multicolumn{2}{l}{$^a$ H$_2$ gas phase reaction energy calculated at PBE-D3/TZVP level}\\
\multicolumn{2}{l}{$^b$ H$_2$ gas phase reaction energy calculated at CCSD(T)/aug-cc-pv5z level}\\
\label{h2_ene}
\end{tabular}
\end{center}
\end{supptable}

\begin{suppfigure*}
\centering
\includegraphics[width=0.9\textwidth]{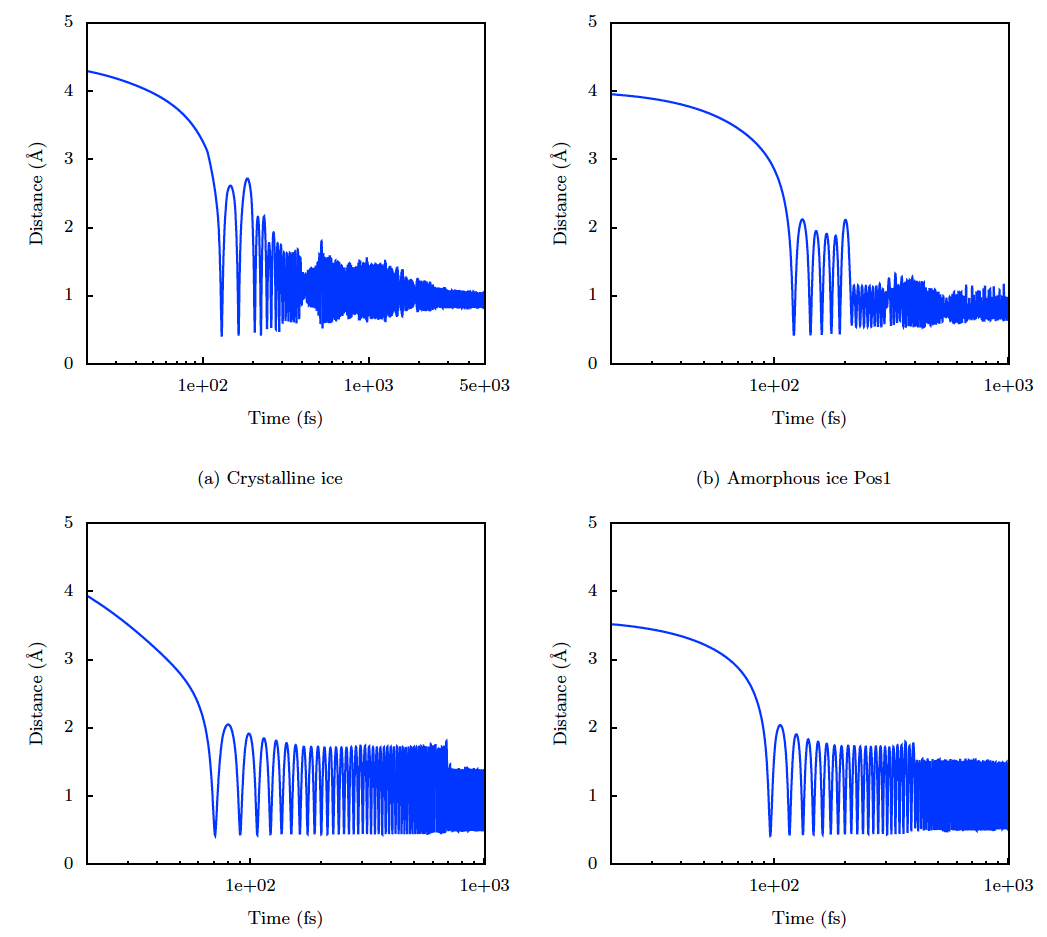}
\caption{Evolution of the H--H bond distance during the AIMD simulations.}
\label{hh_dist}
\end{suppfigure*}

\begin{suppfigure*}
\centering
\includegraphics[width=0.9\textwidth]{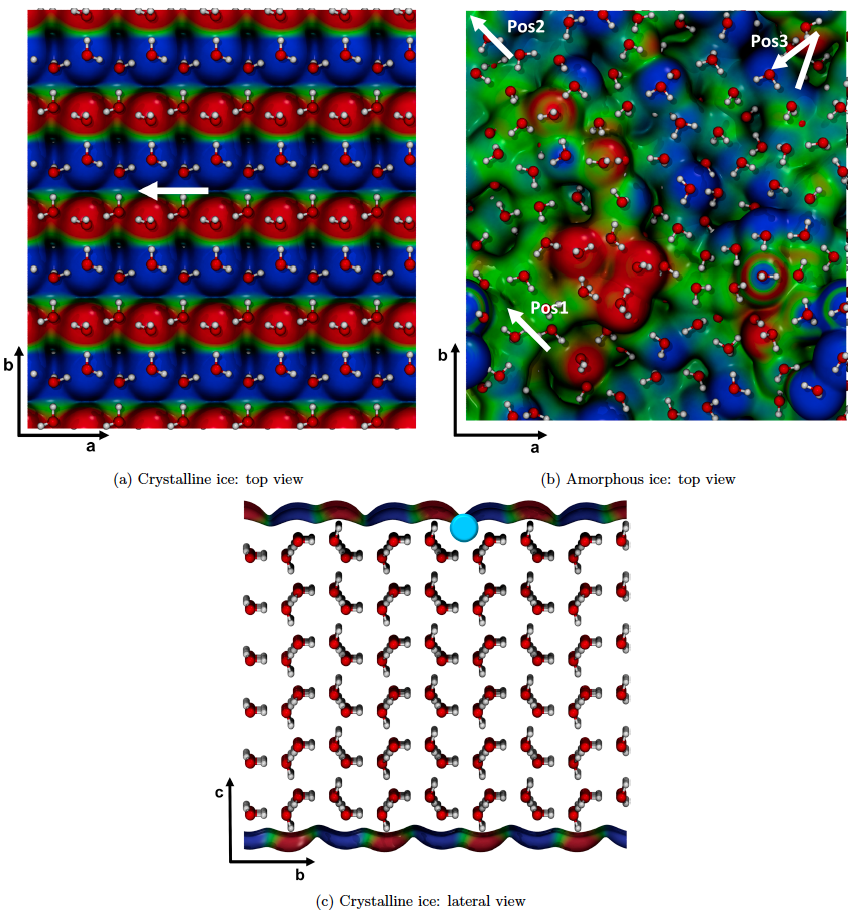}
\caption{Top view of the electrostatic potential map of the crystalline (a) and amorphous (b) ice models. Lateral view of the crystalline ice model (c). White arrows represent the H$_2$ diffusion direction. The light blue circle in the bottom panel represents the cavity inside of which the H$_2$ diffuses over the crystalline ice model. Red and blue zones of the electrostatic maps correspond to positive and negative potentials. O atoms in red, H atoms in white.}
\label{esp}
\end{suppfigure*}

\begin{suppfigure*}
\centering
\includegraphics[width=0.9\textwidth]{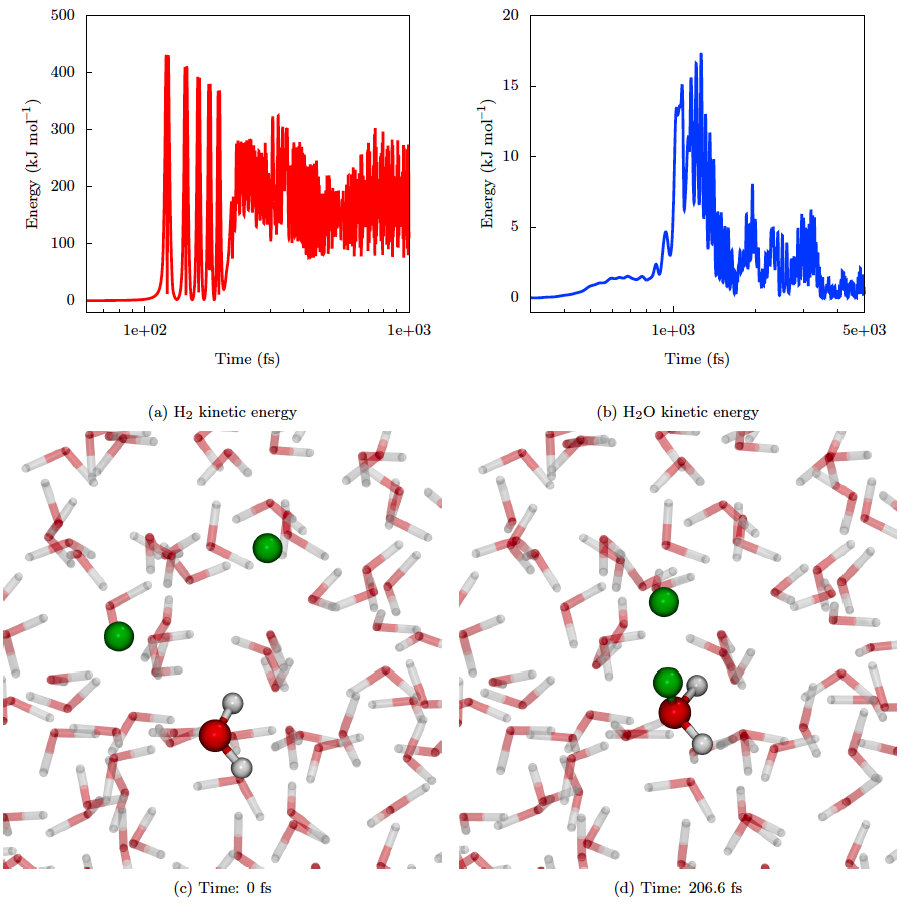}
\caption{Evolution of the kinetic energy of H$_2$ (panel (a)) and a neighbor water molecule (panel (b)) for the amorphous Pos1 case. Starting position of the AIMD simulation (panel (c)) and formation of the H$^{-}$/H$_3$O$^{+}$ adduct after 0.2 picoseconds of simulation (panel (d)). O atoms in red, H atoms in white. The H atoms highlighted in green are the reactants.}
\label{kin_supp}
\end{suppfigure*}


%
\end{document}